\documentclass[aps,prl,twocolumn,amsmath,amssymb]{revtex4}
\usepackage{graphicx}
\usepackage{bm}
\usepackage{color}
\begin{document}

\title{Crystallization of the resonating valence bond liquid as vortex
condensation}

\author{Arnaud Ralko,${^1}$ Michel Ferrero,${^2}$ Federico Becca,${^3}$ 
Dmitri Ivanov,${^4}$ and Fr\'{e}d\'{e}ric Mila${^4}$}
\affiliation{
${^1}$ Laboratoire de Physique Th\'eorique, Universit\'e de Toulouse III,
F-31062 Toulouse, France \\
${^2}$ Centre de Physique Th\'eorique, Ecole Polytechnique, F-91128 Palaiseau 
Cedex, France \\
${^3}$ INFM-Democritos, National Simulation Center and International School 
for Advanced Studies (SISSA), I-34014 Trieste, Italy \\
${^4}$ Institute of Theoretical Physics, Ecole Polytechnique F\'{e}d\'{e}rale 
de Lausanne (EPFL), CH-1015 Lausanne, Switzerland} 

\date{\today}

\begin{abstract}
We show that the liquid-to-crystal quantum phase transition in the
Rokhsar--Kivelson dimer model on the two-dimensional triangular lattice
occurs as a condensation of vortex-like excitations called ``visons''. This
conclusion is drawn from the numerical studies of the vison spectrum in the
liquid phase by using the Green's function Monte Carlo method. We
find that visons remain the lowest excitation throughout the liquid phase
and that their gap decreases continuously to zero at the phase transition.
The nature of the crystal phase and the second order of the phase transition
are in agreement with the earlier prediction of Moessner and Sondhi
[Phys. Rev. B {\bf 63}, 224401 (2001)].
\end{abstract}

\maketitle

The resonating valence bond (RVB) liquid is one of the most intriguing
concepts of today's condensed-matter physics. Conjectured over thirty
years ago for frustrated spin systems,~\cite{fazekas} it still lacks
an adequate quantitative description, and even its emergence in real
physical materials or in realistic spin models is not rigorously established.
In the early days of the RVB paradigm, it has been realized that such
a state should be characterized by a $Z_2$ topological order whose
most prominent consequence is the vortex-like excitation~\cite{chakraborty}
later dubbed ``vison''.~\cite{senthil-vison} 
This type of excitations should be important for the thermodynamics
of RVB spin liquids and serve as a ``smoking gun'' of the RVB phase.
Numerically, low-lying singlet excitations have been found in the Kagome
spin-1/2 Heisenberg antiferromagnet,~\cite{waldtmann} the most promising candidate for the
RVB spin-liquid state, and these singlets have been interpreted as RVB
states in the subspace of nearest-neighbor singlet dimers.~\cite{mila} 
Experimentally, the spin-1/2 Kagome
system with a possible spin-liquid phase has been realized recently
in the ZnCu$_3$(OH)$_6$Cl$_2$ compound.~\cite{kagome-exp} A variational
study of this system with Gutzwiller-projected wave functions also predicts
low-lying excitations of the gauge field [U(1) instead of 
$Z_2$].~\cite{lee-kagome}
Thus the properties of the low-lying collective singlet excitations is one
of the central questions in the study of RVB states.

To model the vison branch of excitations in the RVB state, it has been
proposed to use dimer models instead of spin ones.~\cite{rokhsar} 
In dimer models, the spin degrees of freedom are explicitly frozen, 
and only the vison excitations survive. Regardless of its relation
to microscopic spin models, the quantum dimer model (QDM) of Rokhsar and 
Kivelson (RK) has attracted a lot of attention as a promising way to 
investigate RVB physics, especially on the triangular lattice.
The QDM is defined by:
\begin{eqnarray}\label{hamilt}
H &=& -t \sum
\left(
|\unitlength=1mm
\begin{picture}(6.2,5)
\linethickness{0.5mm}
\put(0.9,-.7){\line(1,2){1.8}}
\put(3.8,-.7){\line(1,2){1.8}}
\end{picture}
\rangle
\langle
\unitlength=1mm
\begin{picture}(6.5,5)
\linethickness{0.3mm}
\put(3.2,2.6){\line(1,0){3.2}}
\put(0.9,-.7){\line(1,0){3.2}}
\end{picture}
|
+h.c.\right) \nonumber \\
&+& \ V \sum \left(
|\unitlength=1mm
\begin{picture}(6.2,5)
\linethickness{0.5mm}
\put(0.9,-.7){\line(1,2){1.8}}
\put(3.8,-.7){\line(1,2){1.8}}
\end{picture}
\rangle
\langle
\unitlength=1mm
\begin{picture}(6.2,5)
\linethickness{0.5mm}
\put(0.9,-.7){\line(1,2){1.8}}
\put(3.8,-.7){\line(1,2){1.8}}
\end{picture}|+
|
\unitlength=1mm
\begin{picture}(6.5,5)
\linethickness{0.3mm}
\put(3.2,2.6){\line(1,0){3.2}}
\put(0.9,-.7){\line(1,0){3.2}}
\end{picture}\rangle
\langle
\begin{picture}(6.5,5)
\linethickness{0.3mm}
\put(3.2,2.6){\line(1,0){3.2}}
\put(0.9,-.7){\line(1,0){3.2}}
\end{picture}
|
\right),
\end{eqnarray}
where the sum runs over all plaquettes (rhombi) including the three possible 
orientations. The kinetic term controlled by the amplitude $t$ flips the
two dimers on every flippable plaquette, i.e., on every plaquette with two
parallel dimers, while the potential term controlled 
by the interaction $V$ describes a repulsion ($V>0$) or an attraction ($V<0$) 
between nearest-neighbor dimers. For dimer models, the situation is much 
more definite than for spin systems: The RVB phase is firmly established and 
studied in detail in the QDM on the triangular lattice.~\cite{moessner}
At the special so-called RK point in the parameter space ($V=t$), 
an analytic proof of the topological degeneracy is 
possible.~\cite{fendley,ioselevich}
At the same RK point, the quantum-mechanical evolution in imaginary time
is equivalent to a classical stochastic process which allows to deduce
the quantum excitation gap from a {\it classical} Monte Carlo
study.~\cite{henley} With such a numerical work, it has been established
that the elementary excitations are indeed visons, that is the gap 
in the vison excitation sector (non-local in terms of dimers) is
lower than that in the dimer sector.~\cite{ivanov-vison}
Dimer excitations are then interpreted as composite two-vison states.

With the Green's function Monte Carlo (GFMC) method,~\cite{nandini,calandra}
we can extend this approach away from the RK point $V=t$ to a wider class of 
systems (provided the ground state is positive definite, i.e., for models
without the sign problem). In application to the QDM, this method allows us 
to study the boundaries of the RVB phase and the transitions to the crystal 
phases. In our earlier works,~\cite{ralko,ralko2} we have computed the static 
and dynamic dimer correlations in the QDM for different values of $V/t$. 
Based on those results, we have refined the phase diagram proposed by 
Moessner and Sondhi on the basis of the quantum Monte Carlo studies 
(at finite temperature).~\cite{moessner}
We have found a crystal phase with a 12-site unit cell
(as predicted in Refs.~\onlinecite{moessner,moessner2}) which destroys
the RVB phase at the critical parameter value $V/t \approx 0.8$.
At the transition point, the static form factor of the crystal decreases
to zero on the crystal side of the transition, while the dimer gap 
also decreases to zero on the liquid side.~\cite{ralko2}

However, an interesting question remains about the fate of visons at this
phase transition. Firstly, visons are expected to be more fundamental
excitations than dimers, at least at the RK point.~\cite{ivanov-vison} 
Secondly, to describe this phase transition, Moessner and Sondhi used the 
mapping of the RK dimer model at $V=0$ onto the frustrated Ising model on the 
dual (hexagonal) lattice with a weak transverse field.~\cite{moessner2} 
In this mapping, the spins of the Ising model correspond to the visons of the 
dimer model and the crystallization transition was described as the ordering 
of Ising spins equivalent to the condensation of visons. They have further 
conjectured that the same type of transition should occur in the dimer model 
as a function of $V/t$.

To test this conjecture, we have performed numerical studies of the dynamic
and static vison correlation functions with the use of the GFMC method.
We find that as the parameter $V/t$ of the QDM decreases below
the RK point, the vison gap gradually decreases and reaches zero
at the presumed crystallization point. On the other side of the transition,
the Bragg peaks in the static vison correlations gradually appear signaling
the crystal phase. This numerical evidence of the crystallization transition
by vison condensation is the main result of our work.

{\it Visons.--} 
Vison operators and corresponding excitations in QDM have been constructed
and considered in detail in 
Refs.~\onlinecite{chakraborty,senthil-vison,ivanov-vison}. Here we briefly
recall the main principles of the construction.

Point-vison operators are defined on the sites of the dual lattice (in our
case, on the triangular plaquettes forming a honeycomb lattice).
In a closed system (without open boundaries), only products of even number
of visons are well defined. The product of two point visons is defined by
\begin{equation}
V_{i} V_{j} = (-1)^\text{no. of dimers intersecting $\Gamma_{ij}$}
\label{vison-definition}
\end{equation}
where $\Gamma_{ij}$ is some path on the dual lattice connecting the
triangular plaquettes $i$ and $j$. With this definition, 
$V_{i} V_{j}$ is independent of the path $\Gamma_{ij}$ up to 
a global sign (the same for all dimer coverings).
Even though only products of even number of visons are defined in this 
way, one can separate the visons far apart in a large system and consider 
them as individual vortex-like excitations.~\cite{chakraborty,senthil-vison} 
Moreover, it has been shown in Ref.~\onlinecite{ivanov-vison} that in a finite 
closed system one can also define a product of vison operators taken at 
different times and hence the correlation function
\begin{equation}
F({\bf r}_{ij},t-t^\prime)=\langle V_i(t)V_j(t^\prime) \rangle \, .
\label{vison-correlation}
\end{equation}
The construction is based on inserting the identity operator 
$V_i(t^\prime)V_i(t^\prime)$ in the product $V_i(t)V_j(t^\prime)$. Then the 
resulting product of four vison operators may be decoupled into the equal-time 
product $V_i(t^\prime)V_j(t^\prime)$ defined by~(\ref{vison-definition}) and 
the equal-position product $V_i(t)V_i(t^\prime)$ which simply counts the 
number of flips between $t$ and $t^\prime$ on the rhombi containing 
$i$.~\cite{ivanov-vison}

Remarkably, the asymptotic behavior of the correlation 
function~(\ref{vison-correlation}) at large imaginary time gives us access to 
the gap in the excitation spectrum for a {\it single} vison, even though the
computation is performed for a finite system.

\begin{figure}
$\begin{array}{cc}
\includegraphics[width=0.2\textwidth,clip]{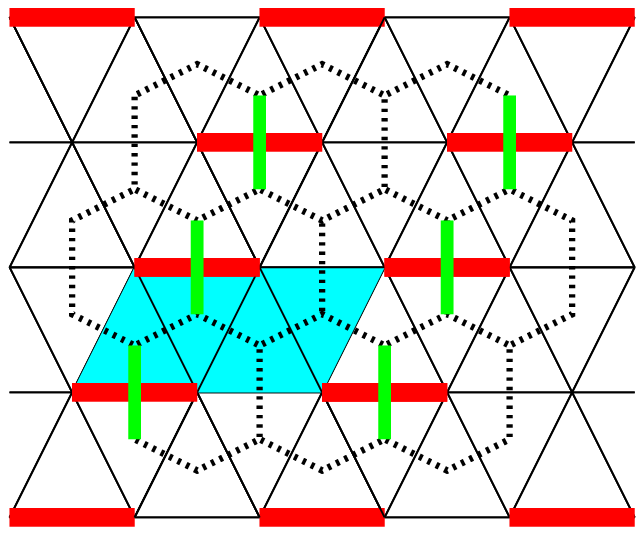} &
\includegraphics[width=0.23\textwidth,clip]{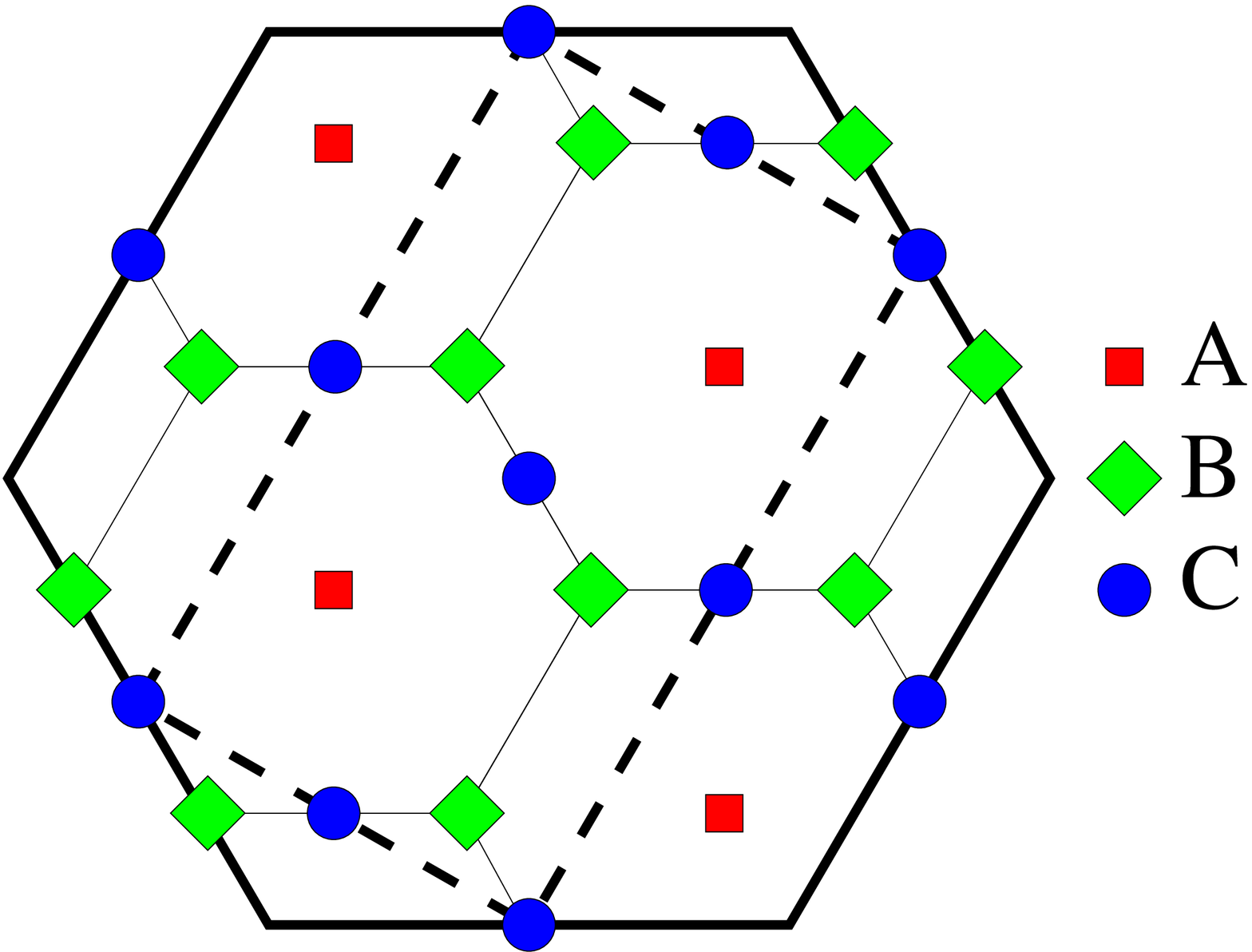}
\end{array}$
\caption{\label{fig:zdb} 
(Color online) Left panel: The reference dimer configuration for fixing
the vison gauge. Dashed black (solid green) lines indicate positive 
(negative) hoppings for
visons on the dual lattice, due to the flux $\pi$ per hexagonal plaquette.
The shaded region is the unit cell for the vison operator
(four sites of the dual lattice). Right panel: Brillouin zones for dimers
and visons. The Brillouin zone of the triangular lattice is the largest 
hexagon; the dashed rectangle denotes the Brillouin zone 
of the vison-like sector. A, B and C denote the high-symmetry points.}
\end{figure}

One subtlety in the vison definition is the necessity of the
gauge fixing and, as a consequence, an artificial doubling of the unit
cell for the correlation function~(\ref{vison-correlation}). The sign
in the vison definition~(\ref{vison-definition}) may be fixed
by choosing the contour $\Gamma_{ij}$ for every point $i$ and a reference
point $j$. If the contour $\Gamma_{ij}$ is deformed to loop around one
site of the lattice, the sign of $V_i V_j$ is reversed, since
there is exactly one dimer starting from that site. As a result, under
the lattice translations, the vison operators $V_i$ transform as particles
moving in a magnetic field with the flux $\pi$ per (hexagonal) plaquette
of the dual lattice. A convenient way to fix a gauge for vison 
operators is to multiply the product of visons~(\ref{vison-definition}) by 
its value in a reference dimer configuration. In terms of the gauge field on 
the dual (hexagonal) lattice, a choice of the reference configuration 
translates into assigning the gauge phase factors of $-1$ to all links 
intersecting the dimers of the reference configuration. The most symmetric 
choice of the reference dimer covering is periodic with two sites of the 
original triangular lattice per unit cell, implying that the unit cell in the
dual lattice contains four sites. Such a reference covering is shown in 
Fig.~\ref{fig:zdb} (left panel). This gauge choice was used in 
Ref.~\onlinecite{ivanov-vison}, and the corresponding Brillouin zone for
vison excitations is shown in the right panel of Fig.~\ref{fig:zdb}.
The high-symmetry points of the vison Brillouin zone are labeled A, B, and C.

\begin{figure}
\vspace{0.5cm}
\includegraphics[width=0.45\textwidth,clip]{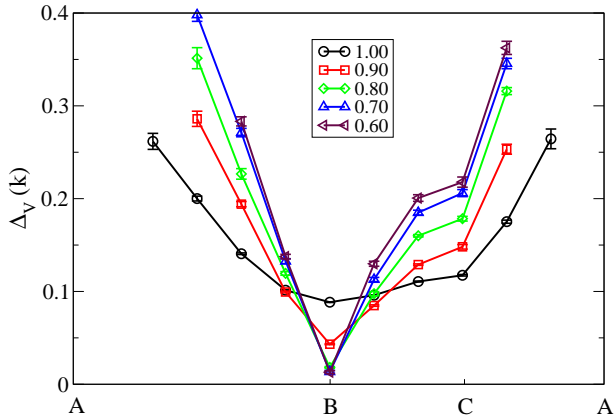}
\caption{\label{fig:visongap_3x12x12} 
(Color online) Lowest-energy excitation spectra for the $432$-site cluster for
different values of $V/t$.}
\end{figure}

{\it Numerical method.--} 
The numerical investigations are based on the Green's function Monte Carlo 
method at zero temperature,~\cite{nandini} adapted to the quantum dimer 
problem~\cite{ralko,ralko2}. With the above prescription, the ground-state 
vison-vison correlation function in imaginary time $F({\bf r}_{ij},\tau)$ 
can be calculated in the same way as the dimer one.~\cite{ralko2}
Note that since the unit cell of the vison operator contains four dual-lattice 
sites, the Fourier transform $F({\bf k},\tau)$ is a $4 \times 4$ matrix for 
each ${\bf k}$ point. Since we are interested in the transition between the 
disordered RVB phase and the $\sqrt{12} \times \sqrt{12}$ crystal, we work 
only with clusters defined by factorization of the infinite triangular 
lattice by the translations ${\bf T}_1 = l {\bf u}_1 + l {\bf u}_2$ and
${\bf T}_2 = -l {\bf u}_1 + 2l {\bf u}_2$. Here ${\bf u}_1=(1,0)$ and 
${\bf u}_2=(1/2,\sqrt{3}/2)$ are the unitary vectors defining the triangular 
lattice and $l$ is an even integer.~\cite{bernu} 
Such clusters contain $N=3 l^2$ sites and can accommodate the 
$\sqrt{12} \times \sqrt{12}$ crystal structure without defects. In the
present work we study numerically clusters up to $l=12$.

\begin{figure}
\vspace{0.5cm}
\includegraphics[width=0.45\textwidth,clip]{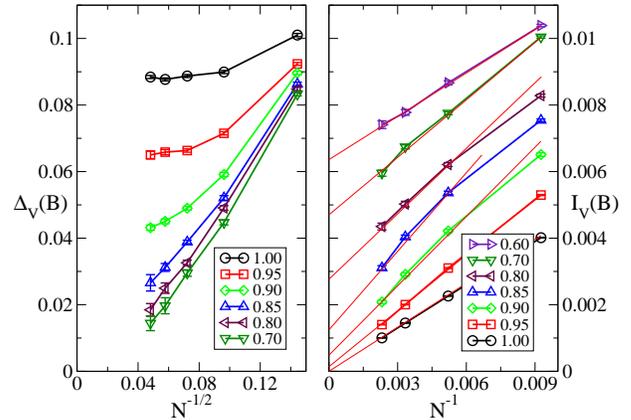}
\caption{\label{fig:sc_B} 
(Color online) Left: Size dependence of the vison gap at point $B$. 
The gap closes at a transition point between $V/t=0.8$ and $V/t=0.85$.
Right: Extrapolation of the Bragg-peak intensity of the vison at point $B$
to the thermodynamic limit.}
\end{figure}

{\it Vison gap.--} 
The vison gap $\Delta_V({\bf k})$, i.e., the energy of the lowest excitation 
with the wave vector ${\bf k}$, can be extracted from the long-time behavior 
of $F({\bf k},\tau)$.~\cite{note} The results for various values of $V/t$ 
obtained on the $3\times 12 \times 12$-site cluster are depicted in 
Fig.~\ref{fig:visongap_3x12x12}. At the RK point, the dispersion reproduces
the results of Ref.~\onlinecite{ivanov-vison}. As $V/t$ decreases below one, 
the shallow minimum at point $B$ deepens, and the gap gradually closes.
To find the value of the vison gap in the thermodynamic limit, we have 
performed a finite-size analysis of $\Delta_V(B)$ shown in Fig.~\ref{fig:sc_B} 
(left panel). We find that $\Delta_V(B)$ extrapolates to a finite value
for $V/t\ge 0.85$ and to zero for $V/t\le 0.8$. 

Similarly to the analysis of the dimer ordering in Ref.~\onlinecite{ralko2},
we have also calculated the intensity of the static vison Bragg peak at
the point B. The size-scaling extrapolation for the Bragg-peak intensity 
is shown in Fig.~\ref{fig:sc_B} (right panel). Near the transition
point, the dependence of the Fourier component of the vison correlation
on the system size is noticeably nonlinear, and the extrapolation has a
low precision. This can be explained by the divergent vison correlation
length at the transition point. We can estimate the vison correlation length
by extrapolating our data in Fig.7 in Ref.~\onlinecite{ralko} to the
approximate transition point. Such an estimate shows that within the 
window $(V/t)_c\pm 0.02$ around the transition point, the correlation
length is of the order or larger than the largest system size considered
in the present work (i.e., $N^{1/2} \sim 20$). Therefore, our results on the
Bragg-peak intensities necessarily lack precision within this window
around the transition point.

\begin{figure}
\includegraphics[width=0.5\textwidth,clip]{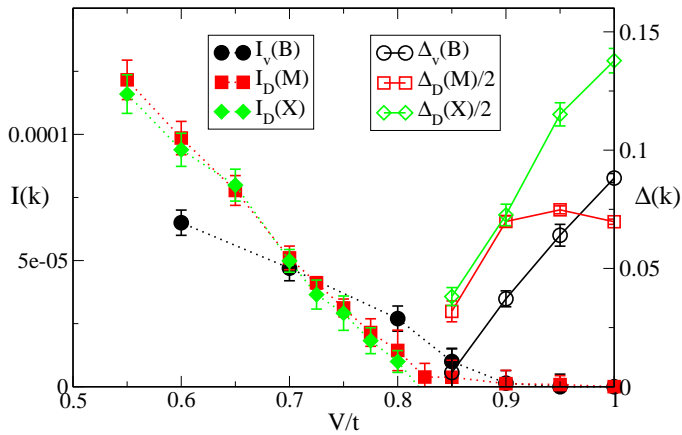}
\caption{\label{fig:gapV} 
(Color online) Dimer gaps at point $X$ and $M$, vison gap at point B,
and the corresponding Bragg peaks as a function of $V/t$. The vison Bragg
peaks are multiplied by 0.01. The dimer gaps are divided by 2.}
\end{figure}

We summarize our results on the vison gap and Bragg peaks
in Fig.~\ref{fig:gapV}, together with our previous data for dimer excitations
from Ref.~\onlinecite{ralko2}. All those data indicate that both vison
and dimer excitations close the gap at the transition point, on the
liquid side. On the crystal side, the Bragg-peak intensity vanishes
continuously, for both dimer and vison correlations. The dimer and
vison data are consistent with each other in locating the transition point,
which may be estimated as $V/t=0.83\pm 0.02$.

Everywhere in the liquid phase, down to the transition point, 
the lowest vison excitation is always below the dimer gap, which
proves that visons are indeed the elementary excitations and allows us
to interpret the transition as a {\it vison condensation}. What is intriguing
though in this respect is the relative magnitudes of the vison gap and of the
dimer gap reported in Ref.~\onlinecite{ralko2}. In the dimer
sector, the lowest excitation should be at most twice the vison gap 
(it could be smaller if there is a bound state). Indeed, the dimer-density 
operator $n_{ij}$ on the bond between two neighboring triangular plaquettes
$i$ and $j$ (which gives 1 when applied to a configuration 
if this bond is occupied by a dimer and 0 otherwise) 
can be written in terms of the vison operators at plaquettes $i$ and $j$ as
\begin{equation}
n_{ij}=\frac{1}{2}(1-V_i V_j)
\label{dimer-two-visons}
\end{equation}
and thus corresponds to two-vison excitations (since point visons do not
produce exact eigenstates, dimer operators also generate higher excitations
with even numbers of visons).
However, at point $X$, the energy of the dimer excitation reported in 
Ref.~\onlinecite{ralko2} is always larger than twice the vison gap. The same
is true at point $M$ for $V/t<0.97$.
To resolve this apparent
inconsistency, we have calculated at the RK point ($V/t=1$)
the dynamics of the product of two visons separated by a distance ${\bf R}$ 
larger than one lattice spacing. The energy of such a composite 
excitation is extracted from the correlation function
$\langle V_i(\tau)V_{i+{\bf R}}(\tau) V_j(0)V_{j+{\bf R}}(0) \rangle$.
Interestingly, the obtained gap at point $X$ {\it decreases} 
upon increasing ${\bf R}$, while at point $M$ the dependence on ${\bf R}$
is the opposite. Both those gaps approach $2\Delta_V(B)$ at large ${\bf R}$.

This suggests the following interpretation: Except close to the RK point
($V/t=1$) and around the $M$ point, where they form a bound state, visons
repel, 
and, therefore, the 
matrix element of the dimer operator between the ground state and the lowest 
state in the two-vison continuum is extremely small. Due to this vison 
repulsion, we are not able to detect the bottom of the two-vison continuum 
from the dimer-dimer correlations. This interpretation further suggests that 
the energy detected with the dimer-dimer correlations in 
Ref.~\onlinecite{ralko2} corresponds not to an eigenstate, but to 
a {\it resonance} inside the two-vison continuum whose lower edge is twice 
the vison gap.

{\it Summary.--}
By using the GFMC method, we have numerically studied the quantum phase
transition from the RVB liquid to a crystal phase in the Rokhsar--Kivelson
dimer model on the triangular lattice. Our results indicate a second-order
phase transition consistent with the conjecture of Moessner and Sondhi based
on a relation to a frustrated Ising model.~\cite{moessner2} At the
crystallization transition, the vison excitation becomes gapless, and
in the crystal phase the vison operator acquires a finite expectation
value. In this respect, it may be suitable to call this crystal order
a ``vison crystal'', since it has a long-range vison correlation 
$\langle V_i V_j \rangle$.

These results definitely establish visons as the relevant excitations to describe
the transition between a valence-bond solid and a resonating valence-bond
liquid. This calls for further investigation of several aspects of their physics,
such as the nature of their mutual interaction and the formation of bound
states. In that respect, it would be extremely useful to come up with a simple
analytical approach, like the Holstein-Primakoff bosonic description of
spin waves. A possibility might be to build further on the 
relationship to the frustrated Ising model in a transverse field.
Work is in progress along these lines.

F.M. is very grateful to Gr\'egoire Misguich for enlightening discussions
on several aspects of this problem. This project has been supported by the 
Swiss National Fund and by MaNEP. F.B. is supported by CNR-INFM.


\begin{thebibliography}{99}
\bibitem{fazekas} P.W. Anderson, Mat. Res. Bull. {\bf 8}, 153 (1973);
   P.Fazekas and P.W. Anderson, Philos. Mag. {\bf 30}, 423 (1974).
\bibitem{chakraborty} N. Read and B. Chakraborty, \prb {\bf 40}, 7133 (1989).
\bibitem{senthil-vison} T. Senthil and M.P.A. Fisher, \prb {\bf 62}, 7850 
   (2000); {\bf 63}, 134521 (2001).
\bibitem{waldtmann} C. Waldtmann {\it et al.}, Eur. Phys. J. B {\bf 2}, 
   501 (1998).
\bibitem{mila} F. Mila, \prl {\bf 81}, 2356 (1998); M. Mambrini and F. Mila, 
{\it Eur. Phys. J. B} {\bf 17}, 651 (2000).
\bibitem{kagome-exp} P. Mendels {\it et al.}, \prl {\bf 98}, 077204 (2007);
   J.S. Helton {\it et al.}, \prl {\bf 98}, 107204 (2007);
   O. Ofer {\it et al.}, cond-mat/0610540.
\bibitem{lee-kagome} Y. Ran, M. Hermele, P.A. Lee, and X.-G. Wen, \prl 
   {\bf 98}, 117205 (2007).
\bibitem{rokhsar} D.S. Rokhsar and S.A. Kivelson, \prl {\bf 61}, 2376 (1988).
\bibitem{moessner} R. Moessner and S.L. Sondhi, \prl {\bf 86}, 1881 (2001).
\bibitem{fendley} P. Fendley, R. Moessner, and S.L. Sondhi, \prb {\bf 66}, 
   214513 (2002).
\bibitem{ioselevich} A. Ioselevich, D.A. Ivanov, and M.V. Feigelman, \prb
   {\bf 66}, 174405 (2002).
\bibitem{henley} C. Henley, cond-mat/9607222; cond-mat/0311345.
\bibitem{ivanov-vison} D. Ivanov, \prb {\bf 70}, 094430 (2004).
\bibitem{nandini} N. Trivedi and D.M. Ceperley, \prb {\bf 41}, 4552 (1990).
\bibitem{calandra} M. Calandra and S. Sorella, \prb {\bf 57}, 11446 (1998).
\bibitem{ralko} A. Ralko, M. Ferrero, F. Becca, D. Ivanov, and F. Mila, \prb 
   {\bf 71}, 224109 (2005).
\bibitem{ralko2} A. Ralko, M. Ferrero, F. Becca, D. Ivanov, and F. Mila, \prb 
   {\bf 74}, 134301 (2006).
\bibitem{moessner2} R. Moessner and S.L. Sondhi, \prb {\bf 63}, 224401 (2001). 
\bibitem{bernu} B. Bernu, P. Lecheminant, C. Lhuillier, and L. Pierre, \prb 
   {\bf 50}, 10048 (1994).
\bibitem{note} In practice, we extract the vison gap from the trace of 
   the $4 \times 4$ matrix $F({\bf k},\tau)$ for each ${\bf k}$ point.
\end{thebibliography}
\end{document}